\documentclass[10pt]{article}
\usepackage{graphicx}
\usepackage{amsmath}
\usepackage{amssymb}
\usepackage{caption2}
\begin{document}
\bibliographystyle{prsty}
\begin{center}
{\large {\bf \sc{  Analysis of the $X(6600)$, $X(6900)$, $X(7300)$ and related tetraquark states with the QCD sum rules }}} \\[2mm]
Zhi-Gang  Wang \footnote{E-mail: zgwang@aliyun.com.  }   \\
 Department of Physics, North China Electric Power University, Baoding 071003, P. R. China
\end{center}

\begin{abstract}
In this work, we re-investigate the mass spectrum of the ground state, first, second and third radial excited states of the diquark-antidiquark type fully-charm  tetraquark states with the QCD sum rules plus Regge trajectories. We take account of the CMS and ATLAS experimental data and preform a self-consistent analysis, then try to make possible assignments of the $X(6600)$, $X(6900)$ and $X(7300)$ in the picture of tetraquark states with the $J^{PC}=0^{++}$ or $1^{+-}$.
\end{abstract}



\section{Introduction}
In  recent years, there have been discovered a number of exotic  charmonium-like and bottomonium-like  states. If they are  genuine resonances, irrespective of the tetraquark states or molecular states, there are two heavy valence quarks and two light valence  quarks. Therefore the dynamics is much complex compared with the configurations consist of four heavy valence quarks,  due to the light degrees of freedom considering the $u$, $d$ and $s$ quarks.

In 2020, the LHCb collaboration reported the evidences of two fully-charm  tetraquark  candidates  in the $J/\psi J/\psi$  mass spectrum  at transverse momentum $p_T> 5.2\,\rm{GeV}$. To achieve those goals, they used proton-proton collision data at centre-of-mass energies of $\sqrt{s}=7$, $8$ and $13\,\rm{TeV}$ recorded by the LHCb experiment, which  correspond  to an integrated luminosity of $9\, \rm{fb}^{-1}$  \cite{LHCb-cccc-2006}. They observed  a broad structure (which maybe consist of several overlapping narrow tetraquark states) above the $J/\psi J/\psi$ threshold ranging from 6.2 to 6.8 GeV and a narrow structure at about 6.9 GeV  with the significance of larger  than $5\sigma$. In addition, they also observed  some vague structures around 7.2 GeV.

Very recently, at the ICHEP 2022 conference, the ATLAS collaboration reported the evidences of several fully-charm tetraquark  excesses
 decaying into a pair of charmonium states in the four $\mu$ final states. To achieve those goals, they used proton-proton collision data at centre-of-mass energies of $\sqrt{s}=13\,\rm{TeV}$ recorded by the ATLAS experiment, which correspond to an integrated luminosity of $139\,\rm{fb}^{-1}$
\cite{Atlas2022}. They observed statistically significant excesses in the $J/\psi J/\psi$ channel, which are consistent with a narrow resonance at about $6.9\,\rm{GeV}$ and a broader structure at much lower mass. And they also observed a statistically significant excess in the $J/\psi \psi^\prime$ channel.

Also at the ICHEP 2022 conference, the CMS collaboration reported their measurements on the $J/\psi J/\psi$ mass spectrum by using proton-proton data at center-of-mass energies of $13\,\rm{ TeV}$, which  correspond to an integrated luminosity of $135$~fb$^{-1}$ \cite{Cms2022}.  They not only confirmed the existence of the $X(6900)$ previously reported by the LHCb collaboration with significance larger than 9.4$\sigma$, but also observed the signals of two new peaking structures. They used the relativistic $S$-wave Breit-Wigner functions to parameterize the resonance structures,  the significances  of the  new structures $X(6600)$ and $X(7300)$ are larger than $5.7\,\sigma$ and $4.1\,\sigma$, respectively. The measured Breit-Wigner masses and widths are \cite{Cms2022},
\begin{eqnarray}
M_{X(6600)}&=&6552\pm10\pm12 ~ \mathrm{MeV},~\nonumber \\
\Gamma_{X(6600)}&=&124 \pm 29\pm 34 ~ \mathrm{MeV}, \nonumber \\
M_{X(6900)}&=&6927\pm9\pm5 ~ \mathrm{MeV},~\nonumber \\
\Gamma_{X(6900)}&=&122 \pm 22\pm 19 ~ \mathrm{MeV}, \nonumber \\
M_{X(7300)}&=&7287\pm19\pm5 ~ \mathrm{MeV},~ \nonumber \\
 \Gamma_{X(7300)}&=&95 \pm 46\pm 20 ~ \mathrm{MeV}. \nonumber
\end{eqnarray}

The predicted masses of the fully-heavy  tetraquark (molecular) states from the  different quark models before and after the LHCb experiment \cite{LHCb-cccc-2006} lie either  above or below the $J/\psi J/\psi$ or $\Upsilon\Upsilon$ threshold, and vary at a large range \cite{Rosner-2017,WZG-QQQQ-1,Chen-2017,Wu-2018,FKGuo-2018-Anwar,Polosa-2018,Hughes-2018,Navarra-2019,Bai-2016,WZG-QQQQ-2,Roberts-2020,
WZG-CPC-cccc,PingJL-2020,Zhong-2019,WangJZ-Produ-mass,Zhong-2021,Zhu-2021-NPB,WZG-IJMPA-cccc,ZhangJR-PRD,QiaoCF-2021,
GuoFK-2021-PRL,XKDong-SB-2021,Gong-Zhong-2022}, none of them are fully consistent with the more precise  measurements of the CMS collaboration \cite{Cms2022}.
Undoubtedly, the new CMS experimental data can provide more refine hints to decode the novel peaking structures appeared in the  $J/\psi J/\psi$ mass spectrum, and  serve as more  powerful  constraints on the theoretical models and maybe shed  light on the nature of  the exotic states.

In Refs.\cite{WZG-QQQQ-1,WZG-QQQQ-2}, we take the axialvector diquark operators in color antitriplet as the elementary building blocks to construct the interpolating currents, and explore the mass spectrum of the ground states of the scalar, axialvector, vector and tensor fully-heavy tetraquark states with the QCD sum rules. Subsequently, we explore the  mass spectrum of the first radial excited  states of the scalar, axialvector, vector and tensor diquark-antidiquark type fully-charm  tetraquark states with the QCD sum rules. Then we  resort to the Regge trajectories to acquire
 the masses of the second radial excited states, and make possible assignments of the LHCb's new structures \cite{WZG-CPC-cccc}. In Ref.\cite{WZG-IJMPA-cccc}, we introduce a relative P-wave to construct the doubly-charm vector diquark operator (not axialvector diquark operator, in Ref.\cite{WZG-IJMPA-cccc}, there is a typo error in the name of the diquark operator). Then we take the vector diquark operator as the basic constituent, and   construct the scalar and tensor local four-quark currents to explore  the scalar, axialvector and tensor fully-charm tetraquark states with the QCD sum rules.

In this work, we re-investigate  the mass spectrum of the ground state, first radial, second radial and third radial excited diquark-antidiquark type fully-charm  tetraquark states with the QCD sum rules. We take account of the new CMS and ATLAS experimental data and preform a self-consistent  analysis, and try to make possible assignments of the $X(6600)$, $X(6900)$ and $X(7300)$ consistently.

The article is arranged as follows:  we acquire the QCD sum rules for the ground states  and  first radial excited states of the  fully-charm  tetraquark states in section 2; in section 3, we present the numerical results and resort to the Regge trajectories to acquire the masses of the second and third  radial excited states; section 4 is reserved for our conclusion.

\section{QCD sum rules for  the ground states and first radial excited states }
Firstly, we  write down  the two-point correlation functions   $\Pi (p)$ and $\Pi_{\mu\nu\alpha\beta}(p)$,
\begin{eqnarray}
\Pi(p)&=&i\int d^4x e^{ip \cdot x} \langle0|T\left\{J(x)J^{\dagger}(0)\right\}|0\rangle \, ,\nonumber\\
\Pi_{\mu\nu\alpha\beta}(p)&=&i\int d^4x e^{ip \cdot x} \langle0|T\left\{J_{\mu\nu}(x)J_{\alpha\beta}^{\dagger}(0)\right\}|0\rangle \, ,
\end{eqnarray}
where $J_{\mu\nu}(x)=J^1_{\mu\nu}(x)$, $J^2_{\mu\nu}(x)$,
\begin{eqnarray}
J(x)&=&\varepsilon^{ijk}\varepsilon^{imn}c^{T}_j(x)C\gamma_\mu c_k(x) \bar{c}_m(x)\gamma^\mu C \bar{c}^{T}_n(x) \, , \nonumber\\
J^1_{\mu\nu}(x)&=&\varepsilon^{ijk}\varepsilon^{imn}\Big\{c^{T}_j(x)C\gamma_\mu c_k(x) \bar{c}_m(x) \gamma_\nu C \bar{c}^{T}_n(x)-c^{T}_j(x)C\gamma_\nu  c_k(x) \bar{c}_m(x)\gamma_\mu C \bar{c}^{T}_n(x) \Big\} \, , \nonumber \\
J^2_{\mu\nu}(x)&=&\frac{\varepsilon^{ijk}\varepsilon^{imn}}{\sqrt{2}}\Big\{c^{T}_j(x)C\gamma_\mu c_k(x) \bar{c}_m(x) \gamma_\nu C \bar{c}^{T}_n(x)+c^{T}_j(x)C\gamma_\nu  c_k(x) \bar{c}_m(x)\gamma_\mu C \bar{c}^{T}_n(x) \Big\} \, , \nonumber \\
\end{eqnarray}
 the $i$, $j$, $k$, $m$, $n$ are color indexes, and the $C$ is the charge conjugation matrix \cite{WZG-QQQQ-1,WZG-QQQQ-2}. We construct  the  local four-quark currents $J(x)$, $J^1_{\mu\nu}(x)$ and $J^2_{\mu\nu}(x)$ to interpolate the   fully-charm  tetraquark states with the quantum numbers $J^{PC}=0^{++}$, $1^{+-}$, $1^{--}$ and $2^{++}$, respectively. The Lorentz indexes $\mu$  and $\nu$ in the current $J^1_{\mu\nu}(x)$ are antisymmetric,  thus it has both the spin-parity $J^P=1^+$ and $1^-$ components. The negative parity indicates that there exists an additional P-wave, which changes the parity, the vector tetraquark states are P-wave states, while the scalar,
 axialvector and tensor tetraquark states are S-wave states.

We take account of the possible  current-hadron couplings, and isolate the ground state contributions of the fully-charm tetraquark states,
\begin{eqnarray}
\Pi (p) &=&\frac{\lambda_X^2}{M^2_X-p^2} +\cdots \, \, , \nonumber\\
&=&\Pi_S(p^2)\, ,
\end{eqnarray}
\begin{eqnarray}
\Pi^1_{\mu\nu\alpha\beta}(p)&=&\frac{\lambda_{ X}^2}{M_{X}^2\left(M_{X}^2-p^2\right)}\left(p^2g_{\mu\alpha}g_{\nu\beta} -p^2g_{\mu\beta}g_{\nu\alpha} -g_{\mu\alpha}p_{\nu}p_{\beta}-g_{\nu\beta}p_{\mu}p_{\alpha}+g_{\mu\beta}p_{\nu}p_{\alpha}+g_{\nu\alpha}p_{\mu}p_{\beta}\right) \nonumber\\
&&+\frac{\lambda_{ Y}^2}{M_{Y}^2\left(M_{Y}^2-p^2\right)}\left( -g_{\mu\alpha}p_{\nu}p_{\beta}-g_{\nu\beta}p_{\mu}p_{\alpha}+g_{\mu\beta}p_{\nu}p_{\alpha}+g_{\nu\alpha}p_{\mu}p_{\beta}\right) +\cdots \, \, ,\nonumber\\
&=&\Pi_{A}(p^2)\left(p^2g_{\mu\alpha}g_{\nu\beta} -p^2g_{\mu\beta}g_{\nu\alpha} -g_{\mu\alpha}p_{\nu}p_{\beta}-g_{\nu\beta}p_{\mu}p_{\alpha}+g_{\mu\beta}p_{\nu}p_{\alpha}+g_{\nu\alpha}p_{\mu}p_{\beta}\right) \nonumber\\
&&+\Pi_{V}(p^2)\left( -g_{\mu\alpha}p_{\nu}p_{\beta}-g_{\nu\beta}p_{\mu}p_{\alpha}+g_{\mu\beta}p_{\nu}p_{\alpha}+g_{\nu\alpha}p_{\mu}p_{\beta}\right) \, .
\end{eqnarray}
\begin{eqnarray}
\Pi^2_{\mu\nu\alpha\beta} (p) &=&\frac{\lambda_X^2}{M_X^2-p^2}\left( \frac{\widetilde{g}_{\mu\alpha}\widetilde{g}_{\nu\beta}+\widetilde{g}_{\mu\beta}\widetilde{g}_{\nu\alpha}}{2}-\frac{\widetilde{g}_{\mu\nu}\widetilde{g}_{\alpha\beta}}{3}\right) +\cdots \, \, ,  \nonumber\\
&=&\Pi_{T}(p^2)\left( \frac{\widetilde{g}_{\mu\alpha}\widetilde{g}_{\nu\beta}+\widetilde{g}_{\mu\beta}\widetilde{g}_{\nu\alpha}}{2}-\frac{\widetilde{g}_{\mu\nu}\widetilde{g}_{\alpha\beta}}{3}\right) +\cdots \, \, ,
\end{eqnarray}
where $\widetilde{g}_{\mu\nu}=g_{\mu\nu}-\frac{p_{\mu}p_{\nu}}{p^2}$, the pole residues  $\lambda_{X}$ and $\lambda_{Y}$ are defined by
\begin{eqnarray}
 \langle 0|J (0)|X (p)\rangle &=& \lambda_{X}     \, , \nonumber\\
  \langle 0|J^1_{\mu\nu}(0)|X(p)\rangle &=& \frac{\lambda_{X}}{M_{X}} \, \varepsilon_{\mu\nu\alpha\beta} \, \varepsilon^{\alpha}p^{\beta}\, , \nonumber\\
 \langle 0|J^1_{\mu\nu}(0)|Y(p)\rangle &=& \frac{\lambda_{Y}}{M_{Y}} \left(\varepsilon_{\mu}p_{\nu}-\varepsilon_{\nu}p_{\mu} \right)\, ,\nonumber\\
  \langle 0|J^2_{\mu\nu}(0)|X (p)\rangle &=& \lambda_{X} \, \varepsilon_{\mu\nu}   \, ,
\end{eqnarray}
the $\varepsilon_{\mu}$ and $\varepsilon_{\mu\nu} $ are the  polarization vectors of the tetraquark states with the spin $J=1$ and $2$, respectively. We add the subscripts $S$, $A$, $V$ and $T$ to represent the scalar (S), axialvector (A), vector (V) and tensor (T) tetraquark states, respectively.

If we isolate the ground state plus the first radial excited state in all the channels, we acquire the hadron representation,
\begin{eqnarray}
\Pi_{S/T}(p^2)&=&\frac{\lambda_X^2}{M^2_X-p^2} +\frac{\lambda_{X^\prime}^2}{M^2_{X^\prime}-p^2}+\cdots \, , \nonumber\\
\Pi_{A/V}(p^2)&=&\frac{\lambda_{ X/Y}^2}{M_{X/Y}^2\left(M_{X/Y}^2-p^2\right)}+\frac{\lambda_{ X^{\prime}/Y^{\prime}}^2}{M_{X^{\prime}/Y^{\prime}}^2\left(M_{X^{\prime}/Y^{\prime}}^2-p^2\right)}+\cdots\, .
\end{eqnarray}
We introduce the tensor operators $P_{A}^{\mu\nu\alpha\beta}$ and $P_{V}^{\mu\nu\alpha\beta}$ and project out the axialvector and vector components $\Pi_{A}(p^2)$ and $\Pi_{V}(p^2)$ unambiguously,
\begin{eqnarray}
\widetilde{\Pi}_{A}(p^2)&=&p^2\Pi_{A}(p^2)=P_{A}^{\mu\nu\alpha\beta}\Pi_{\mu\nu\alpha\beta}(p) \, , \nonumber\\
\widetilde{\Pi}_{V}(p^2)&=&p^2\Pi_{V}(p^2)=P_{V}^{\mu\nu\alpha\beta}\Pi_{\mu\nu\alpha\beta}(p) \, ,
\end{eqnarray}
where
\begin{eqnarray}
P_{A}^{\mu\nu\alpha\beta}&=&\frac{1}{6}\left( g^{\mu\alpha}-\frac{p^\mu p^\alpha}{p^2}\right)\left( g^{\nu\beta}-\frac{p^\nu p^\beta}{p^2}\right)\, , \nonumber\\
P_{V}^{\mu\nu\alpha\beta}&=&\frac{1}{6}\left( g^{\mu\alpha}-\frac{p^\mu p^\alpha}{p^2}\right)\left( g^{\nu\beta}-\frac{p^\nu p^\beta}{p^2}\right)-\frac{1}{6}g^{\mu\alpha}g^{\nu\beta}\, .
\end{eqnarray}

We accomplish the operator product expansion and acquire  the QCD spectral densities through dispersion relation \cite{WZG-QQQQ-1,WZG-QQQQ-2},
\begin{eqnarray}
\Pi_{S/T}(p^2)&=& \int_{16m_c^2}^{\infty}ds \frac{\rho_{S/T}(s)}{s-p^2}\, ,\nonumber\\
\widetilde{\Pi}_{A/V}(p^2)&=& \int_{16m_c^2}^{\infty}ds \frac{\rho_{A/V}(s)}{s-p^2}\, ,
\end{eqnarray}
where
\begin{eqnarray}
\rho_{S/T}(s)&=&\frac{{\rm Im}\Pi_{S/T}(s)}{\pi}\, , \nonumber\\
\rho_{A/V}(s)&=&\frac{{\rm Im}\widetilde{\Pi}_{A/V}(s)}{\pi}\, ,
\end{eqnarray}
the explicit expressions of the QCD spectral densities can be found in Refs.\cite{WZG-QQQQ-1,WZG-QQQQ-2}.

 We  take the quark-hadron duality below the continuum thresholds  $s_0$ and $s_0^\prime$, respectively,  and accomplish  Borel transform  in regard  to
the variable $P^2=-p^2$ to acquire   the two QCD sum rules:
\begin{eqnarray}\label{QCDST-1S}
\lambda^2_{X/Y}\, \exp\left(-\frac{M^2_{X/Y}}{T^2}\right)&=& \int_{16m_c^2}^{s_0} ds  \rho(s)  \exp\left(-\frac{s}{T^2}\right) \, ,
\end{eqnarray}
\begin{eqnarray}\label{QCDST-2S}
\lambda^2_{X/Y}\, \exp\left(-\frac{M^2_{X/Y}}{T^2}\right)+\lambda^2_{X^\prime/Y^\prime}\, \exp\left(-\frac{M^2_{X^\prime/Y^\prime}}{T^2}\right)&=& \int_{16m_c^2}^{s^\prime_0} ds  \rho(s) \exp\left(-\frac{s}{T^2}\right) \, ,
\end{eqnarray}
where   $\rho(s) =\rho_S(s) $, $\rho_A(s) $, $\rho_V(s)$ and $\rho_T(s)$, the explicit expressions are given in the Appendix.
 We   introduce the symbols $\tau=\frac{1}{T^2}$, $D^n=\left( -\frac{d}{d\tau}\right)^n$, and  take  the subscripts (or the radial quantum numbers) $1$ and $2$ to represent  the  $X$, $Y$ and  $X^\prime$, $Y^\prime$ respectively to acquire more concise expressions.
 Now we rewrite the two QCD sum rules in Eqs.\eqref{QCDST-1S}-\eqref{QCDST-2S} as
\begin{eqnarray}\label{QCDSR-I}
\lambda_1^2\exp\left(-\tau M_1^2 \right)&=&\Pi_{QCD}(\tau) \, ,
\end{eqnarray}
\begin{eqnarray}\label{QCDSR-II-re}
\lambda_1^2\exp\left(-\tau M_1^2 \right)+\lambda_2^2\exp\left(-\tau M_2^2 \right)&=&\Pi^{\prime}_{QCD}(\tau) \, ,
\end{eqnarray}
where we add the subscript $QCD$ to represent the QCD representations  below the continuum thresholds $s_0$ and $s_0^\prime$. We differentiate the QCD sum rules in Eq.\eqref{QCDSR-I} in regard  to $\tau$ to get
the masses of the ground states,
\begin{eqnarray}\label{QCDSR-I-Dr}
M_1^2&=&\frac{D\Pi_{QCD}(\tau)}{\Pi_{QCD}(\tau)}\, .
\end{eqnarray}
We acquire the ground state masses and pole residues according to two coupled QCD sum rules shown in Eq.\eqref{QCDSR-I} and Eq.\eqref{QCDSR-I-Dr} \cite{WZG-QQQQ-1,WZG-QQQQ-2}.

Then we  differentiate  the QCD sum rules in Eq.\eqref{QCDSR-II-re} in regard  to $\tau$ to acquire
\begin{eqnarray}\label{QCDSR-II-Dr}
\lambda_1^2M_1^2\exp\left(-\tau M_1^2 \right)+\lambda_2^2M_2^2\exp\left(-\tau M_2^2 \right)&=&D\Pi^{\prime}_{QCD}(\tau) \, .
\end{eqnarray}
From Eq.\eqref{QCDSR-II-re} and Eq.\eqref{QCDSR-II-Dr}, we can acquire the QCD sum rules,
\begin{eqnarray}\label{QCDSR-II-Residue}
\lambda_i^2\exp\left(-\tau M_i^2 \right)&=&\frac{\left(D-M_j^2\right)\Pi^{\prime}_{QCD}(\tau)}{M_i^2-M_j^2} \, ,
\end{eqnarray}
where  $i \neq j$.
Then we differentiate  the QCD sum rules in Eq.\eqref{QCDSR-II-Residue} in regard  to $\tau$ to get
\begin{eqnarray}
M_i^2&=&\frac{\left(D^2-M_j^2D\right)\Pi_{QCD}^{\prime}(\tau)}{\left(D-M_j^2\right)\Pi_{QCD}^{\prime}(\tau)} \, , \nonumber\\
M_i^4&=&\frac{\left(D^3-M_j^2D^2\right)\Pi_{QCD}^{\prime}(\tau)}{\left(D-M_j^2\right)\Pi_{QCD}^{\prime}(\tau)}\, .
\end{eqnarray}
 The squared masses $M_i^2$ satisfy the  equation,
\begin{eqnarray}
M_i^4-b M_i^2+c&=&0\, ,
\end{eqnarray}
where
\begin{eqnarray}
b&=&\frac{D^3\otimes D^0-D^2\otimes D}{D^2\otimes D^0-D\otimes D}\, , \nonumber\\
c&=&\frac{D^3\otimes D-D^2\otimes D^2}{D^2\otimes D^0-D\otimes D}\, , \nonumber\\
D^j \otimes D^k&=&D^j\Pi^{\prime}_{QCD}(\tau) \,  D^k\Pi^{\prime}_{QCD}(\tau)\, ,
\end{eqnarray}
the subscripts $i=1,2$ and the superscripts  $j,k=0,1,2,3$.
Finally we solve the equation analytically to acquire two solutions, i.e. the masses of the ground states and first radial excited states \cite{Baxi-G,WangZG-4430-1,WangZG-4430-2},
\begin{eqnarray}\label{QCDSR-II-M1}
M_1^2&=&\frac{b-\sqrt{b^2-4c} }{2} \, ,
\end{eqnarray}
\begin{eqnarray}\label{QCDSR-II-M2}
M_2^2&=&\frac{b+\sqrt{b^2-4c} }{2} \, .
\end{eqnarray}
In general, we can acquire the ground state masses either from the QCD sum rules in Eq.\eqref{QCDSR-I-Dr} or in Eq.\eqref{QCDSR-II-M1}, and we prefer the QCD sum rules in Eq.\eqref{QCDSR-I-Dr} considering the larger ground state contributions and less uncertainties from the continuum threshold parameters. We acquire  the masses and pole residues of the first radial excited states from the two coupled QCD sum rules in Eq.\eqref{QCDSR-II-Residue} and in Eq.\eqref{QCDSR-II-M2}.

\section{Numerical results and discussions}

We choose  the traditional  value of the gluon condensate $\langle \frac{\alpha_s
GG}{\pi}\rangle=0.012\pm0.004\,\rm{GeV}^4$
\cite{SVZ79-1,SVZ79-2,Reinders85,ColangeloReview}, and  take the $\overline{MS}$ (modified minimal subtraction
scheme) mass $m_{c}(m_c)=(1.275\pm0.025)\,\rm{GeV}$
 from the Particle Data Group \cite{PDG}.
We take  account of
the energy-scale dependence of  the  $\overline{MS}$  mass from the renormalization group equation,
 \begin{eqnarray}\label{MSmc}
m_c(\mu)&=&m_c(m_c)\left[\frac{\alpha_{s}(\mu)}{\alpha_{s}(m_c)}\right]^{\frac{12}{25}} \, ,\nonumber\\
\alpha_s(\mu)&=&\frac{1}{b_0t}\left[1-\frac{b_1}{b_0^2}\frac{\log t}{t} +\frac{b_1^2(\log^2{t}-\log{t}-1)+b_0b_2}{b_0^4t^2}\right]\, ,
\end{eqnarray}
  where $t=\log \frac{\mu^2}{\Lambda^2}$, $b_0=\frac{33-2n_f}{12\pi}$, $b_1=\frac{153-19n_f}{24\pi^2}$, $b_2=\frac{2857-\frac{5033}{9}n_f+\frac{325}{27}n_f^2}{128\pi^3}$,  $\Lambda=213\,\rm{MeV}$, $296\,\rm{MeV}$  and  $339\,\rm{MeV}$ for the flavors  $n_f=5$, $4$ and $3$, respectively  \cite{PDG}. We explore the properties of the fully-charm tetraquark states and  take the flavor number $n_f=4$.

The values of the $\overline{MS}$ mass of the $c$-quark listed in { \it  The Review of Particle Physics} in 2012, 2014, 2016, (2017) 2018 and 2020   were $m_{c}(m_c)=1.275\pm0.025\,\rm{GeV}$, $1.275\pm0.025\,\rm{GeV}$, $1.27 \pm 0.03\,\rm{GeV}$, ($1.28\pm0.03\,\rm{GeV}$) $1.275^{+0.025}_{-0.035}\,\rm{GeV}$ and $1.27 \pm 0.02\,\rm{GeV}$, respectively. In the QCD sum rules for the hidden-charm tetraquark (molecular) states, we usually choose the value  $1.275\pm0.025\,\rm{GeV}$ from  { \it  The Review of Particle Physics (2012)}, and adopted  the value  ever since \cite{WZG-formula-mole,WZG-formula,WZG-PRD-hidden-charm,WZG-CPC-Zcs3985,WZG-P4312-IJMPA,WZG-Pcs4459-IJMPA}. In Ref.\cite{WZG-QQQQ-1}, we choose the value $m_c(m_c)=1.275\pm0.025\,\rm{GeV}$  for the $c$-quark $\overline{MS}$ mass, while in Ref.\cite{WZG-QQQQ-2}, we choose the value  $m_c(m_c)=1.28\pm0.03\,\rm{GeV}$, in the present work, we choose the uniform value $m_c(m_c)=1.275\pm0.025\,\rm{GeV}$, just like in our previous works \cite{WZG-formula-mole,WZG-formula,WZG-PRD-hidden-charm,WZG-CPC-Zcs3985,WZG-P4312-IJMPA,WZG-Pcs4459-IJMPA}, and perform a updated analysis.

We should choose the suitable energy scales $\mu$  (therefore the suitable charm quark mass $m_c(\mu)$ according to Eq.\eqref{MSmc}) and continuum threshold parameters $s_0$ and  $s^\prime_0$ to acquire very flat platforms with variations of the Borel parameters. Secondly, we should take account of the ground state (plus the first radial excited  state) contributions fully and avoid contaminations from the higher resonances and continuum states. Thirdly, the masses of the ground states and first/second/third radial excited states should obey the Regge trajectories,  and the continuum threshold parameters $s_0$ and $s_0^\prime$ should have no  contradictions with the $M(\rm{2S/2P})$ and $M(\rm{3S/3P})$, respectively. While in Refs.\cite{WZG-QQQQ-1,WZG-QQQQ-2,WZG-CPC-cccc},
the relations among  the $s_0$, $s_0^\prime$, $M(\rm{2S/2P})$ and $M(\rm{3S/3P})$ cannot exclude all possible contaminations.

 In Ref.\cite{WZG-QQQQ-1}, we observe that the predicted tetraquark masses decease with the increase of the energy scales of the QCD spectral densities, only suitable energy scales can lead to very flat Borel platforms, too large or too small energy scales fail to work.
 For example,  the energy scale $\mu=1.4-2.0\,\rm{GeV}$ works for the scalar and tensor fully-charm tetraquark states, see the Fig.1 in Ref.\cite{WZG-QQQQ-1}. And we choose the largest energy scale $\mu=2.0\,\rm{GeV}$, which leads to the predicted scalar tetraquark mass $M_S=5.99\,{\rm{GeV}}\,({\rm central\, value})< 2M_{J/\psi}$.

  In the picture of tetraquark states, the $Z_c(3900)$ ($Z_{cs}(3985)$) and $Z_c(4020)$ can be  tentatively assigned to be  the $S\bar{A}-A\bar{S}$ and $A\bar{A}$ type tetraquark states, respectively, according to the calculations based on the QCD sum rules \cite{WZG-PRD-hidden-charm,WZG-CPC-Zcs3985}. And they lie nearby the $D\bar{D}^*$ ($D\bar{D}_s^*$ and $D_s\bar{D}^*$) and $D^*\bar{D}^*$ thresholds, respectively.
   On the other hand, in the picture of diquark-diquark-antiquark type pentaquark states,
 we can reproduce the masses of the  $P_c(4312)$, $P_c(4380)$, $P_c(4440)$, $P_c(4457)$ and $P_{cs}(4459)$ states according to the calculations based on the QCD sum rules  \cite{WZG-P4312-IJMPA,WZG-Pcs4459-IJMPA}. And they lie nearby  the charmed  meson-baryon pairs $\bar{D}\Sigma_c$, $\bar{D}\Sigma_c^*$, $\bar{D}^*\Sigma_c$, $\bar{D}^*\Sigma_c$ and $\bar{D}^*\Xi_c^\prime$,  respectively. If the same mechanism holds, the ground state masses of the fully-charm tetraquark states would have the masses about $2M_{J/\psi}$, the energy scale $\mu=2.0\,\rm{GeV}$ is somewhat larger as it leads to smaller tetraquark mass than $2M_{J/\psi}$, and we obtain the lowest tetraquark masses in previous works.

 We search for the suitable energy scales, the Borel parameters, the continuum threshold parameters via trial and error,    and  reach the satisfactory results, which are shown in Tables \ref{mass-residue-1S}-\ref{mass-residue-2S}. From the Tables, we can see clearly that the uniform pole contributions are $(40-60)\%$ and $(60-75)\%$ for the ${\rm 1S/1P}$ states and $\rm{1S+2S/1P+2P}$ states, respectively,  the pole dominance at the hadron side is satisfied very good. In the Borel windows, the dominant contributions come from the perturbative terms,  the operator product expansion  converges very good.

 In calculations, we observe that only in some special intervals of the energy scales we can acquire very flat Borel platforms, and the predicted tetraquark masses decrease with increase of the energy scales. In this work, we choose the lowest energy scales for the ground states and first radial excited states, and try to obtain larger masses (compared with that in our previous works \cite{WZG-QQQQ-1,WZG-QQQQ-2,WZG-CPC-cccc}) to match with the experimental data. In fact, the largest energy scales in those special intervals are not necessary to the
 best energy scales, and not necessary to result in satisfactory predictions to match with the experimental data.

\begin{figure}
 \centering
 \includegraphics[totalheight=6cm,width=7cm]{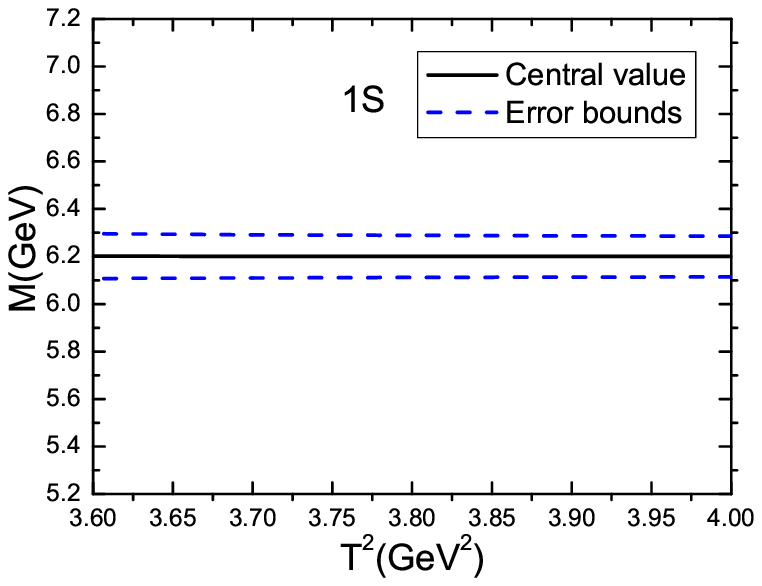}
 \includegraphics[totalheight=6cm,width=7cm]{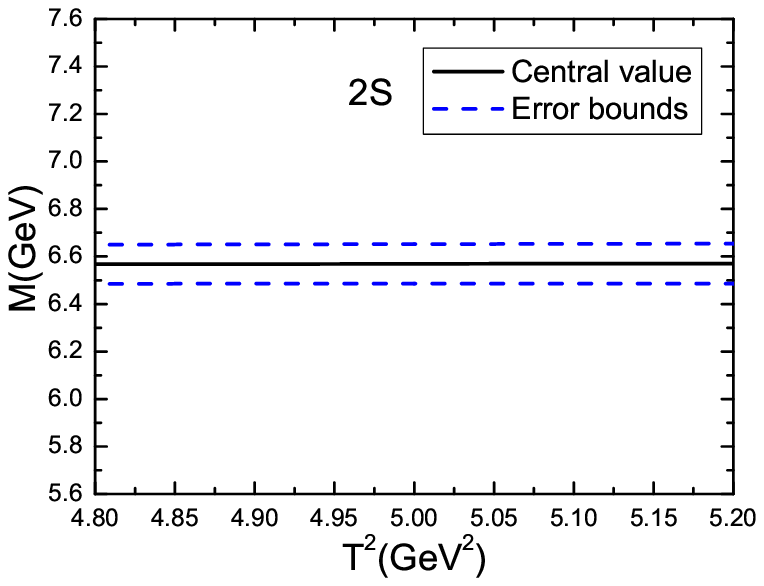}
 \caption{ The masses  of the  fully-charm scalar tetraquark states  with variations of the Borel parameter $T^2$,  where the 1S and 2S represent the ground state and first radial excited state, respectively.  }\label{mass-cccc-S}
\end{figure}

Now we  take account of all uncertainties of the input parameters, and acquire the values of the masses and pole residues of the ground states and first radial excited states of the fully-charm tetraquark states, which are also shown explicitly in Tables  \ref{mass-residue-1S}-\ref{mass-residue-2S}. The predicted masses and pole residues are rather stable with variations of the Borel parameters, the uncertainties come  from the Borel parameters  are very small. For example, in Fig.\ref{mass-cccc-S}, we plot the predicted masses of the 1S and 2S states of the scalar tetraquark states in the Borel windows.

\begin{table}
\begin{center}
\begin{tabular}{|c|c|c|c|c|c|c|c|}\hline\hline
$J^{PC}$        &$T^2(\rm{GeV}^2)$    &$\sqrt{s_0}(\rm{GeV})$ &$\mu(\rm{GeV})$ &pole       &$M_{X/Y}(\rm{GeV})$ &$\lambda_{X/Y}(10^{-1}\rm{GeV}^5)$ \\ \hline

$0^{++}(\rm 1S)$  &$3.6-4.0$          &$6.55\pm0.10$          &$1.4$           &$(39-61)\%$       &$6.20\pm0.10$        &$2.68\pm0.57$  \\ \hline

$1^{+-}(\rm 1S)$  &$3.8-4.2$          &$6.60\pm0.10$          &$1.4$           &$(40-61)\%$       &$6.24\pm0.10$        &$2.18\pm0.44$  \\ \hline

$2^{++}(\rm 1S)$  &$4.0-4.4$          &$6.65\pm0.10$          &$1.4$           &$(39-60)\%$       &$6.27\pm0.09$        &$2.35\pm0.46$  \\ \hline

$1^{--}(\rm 1P)$  &$3.2-3.6$          &$6.70\pm0.10$          &$1.2$           &$(39-63)\%$       &$6.33\pm0.10$        &$0.86\pm0.24$  \\ \hline
\hline

\end{tabular}
\end{center}
\caption{ The Borel parameters, continuum threshold parameters, energy scales,  pole contributions, masses and pole residues of the ground state tetraquark states. }\label{mass-residue-1S}
\end{table}

\begin{table}
\begin{center}
\begin{tabular}{|c|c|c|c|c|c|c|c|}\hline\hline
$J^{PC}$        &$T^2(\rm{GeV}^2)$    &$\sqrt{s_0^\prime}(\rm{GeV})$ &$\mu(\rm{GeV})$ &pole       &$M_{X/Y}(\rm{GeV})$ &$\lambda_{X/Y}(10^{-1}\rm{GeV}^5)$ \\ \hline

$0^{++}(\rm 2S)$  &$4.8-5.2$          &$6.90\pm0.10$          &$2.4$           &$(61-75)\%$       &$6.57\pm0.09$        &$8.12\pm1.21$  \\ \hline

$1^{+-}(\rm 2S)$  &$5.2-5.6$          &$7.00\pm0.10$          &$2.4$           &$(61-75)\%$       &$6.64\pm0.09$        &$6.43\pm0.88$  \\ \hline

$2^{++}(\rm 2S)$  &$5.3-5.7$          &$7.05\pm0.10$          &$2.4$           &$(62-75)\%$       &$6.69\pm0.09$        &$6.84\pm0.92$  \\ \hline

$1^{--}(\rm 2P)$  &$5.0-5.4$          &$7.10\pm0.10$          &$2.4$           &$(60-74)\%$       &$6.74\pm0.09$        &$4.74\pm0.71$  \\ \hline
\hline

\end{tabular}
\end{center}
\caption{ The Borel parameters, continuum threshold parameters, energy scales,  pole contributions, masses and pole residues of the first radial excited  tetraquark states. }\label{mass-residue-2S}
\end{table}

  If the masses of the ground states, the first radial excited states, the second  radial excited states, the third radial excited states, etc. satisfy the  Regge trajectories,
 \begin{eqnarray}
 M_n^2&=&\alpha (n-1)+\alpha_0\, ,
 \end{eqnarray}
  just like the conventional mesons and baryons, where the $\alpha$ and $\alpha_0$ are constants. We use the masses of the ground states and first radial excited states presented  in Table \ref{mass-cccc-Regge}  to fit the  $\alpha$ and $\alpha_0$, then we acquire the masses of the second and third radial excited states, which are also shown in Table \ref{mass-cccc-Regge}. In Table \ref{mass-cccc-Regge}, the upper (lower) bounds in the uncertainties  correspond to the upper (lower) bounds  of the masses in the Regge trajectories one by one.
 From the Tables \ref{mass-residue-1S}-\ref{mass-cccc-Regge}, we can see clearly that the central values, lower bounds and  upper bounds of the masses and the continuum threshold parameters satisfy the relations,
 \begin{eqnarray}
 M_1<\sqrt{s_0}<M_2<\sqrt{s_0^\prime}<M_3\, ,
  \end{eqnarray}
respectively,  there are no contaminations come from the higher resonances and continuum states, while in previous works, there are some contaminations \cite{WZG-QQQQ-1,WZG-QQQQ-2,WZG-CPC-cccc}.   The present calculations are self-consistent.

\begin{table}
\begin{center}
\begin{tabular}{|c|c|c|c|c|c|c|c|}\hline\hline
$J^{PC}$         &$M_{1}(\rm{GeV})$      &$M_{2}(\rm{GeV})$     &$M_{3}(\rm{GeV})$   &$M_{ 4}(\rm{GeV})$   \\ \hline

$0^{++} $        &$6.20\pm0.10$          &$6.57\pm0.09$         &$6.92\pm0.09$       &$7.25\pm0.09$       \\

                 &? $X(6220)$            &? $X(6600/6620)$      &? $X(6900)$         &? $X(7220/7300)$      \\ \hline

$1^{+-} $        &$6.24\pm0.10$          &$6.64\pm0.09$         &$7.03\pm0.09$       &$7.40\pm0.09$      \\

                 &? $X(6220)$            &? $X(6600/6620)$      &                    &         \\ \hline

$2^{++} $        &$6.27\pm0.09$          &$6.69\pm0.09$         &$7.09\pm0.09$       &$7.46\pm0.09$          \\ \hline

$1^{--}$         &$6.33\pm0.10$          &$6.74\pm0.09$         &$7.13\pm0.09$       &$7.50\pm0.09$           \\ \hline
\hline

\end{tabular}
\end{center}
\caption{ The  masses of the fully-charm  tetraquark states with the radial quantum numbers $n=1$, $2$, $3$ and $4$. In the lower lines, we present the  possible assignments.  }\label{mass-cccc-Regge}
\end{table}

From Table \ref{mass-cccc-Regge}, we can see clearly that the predicted masses $M=6.20\pm0.10\,\rm{GeV}$ and $6.24\pm0.10\,\rm{GeV}$ for the 1S tetraquark states with the $J^{PC}=0^{++}$ and $1^{+-}$, respectively, are compatible with the value $6.22\pm0.05 {}^{+0.04}_{-0.05}\,\rm{GeV}$ from the ATLAS collaboration \cite{Atlas2022}. And they  support assigning the $X(6220)$ to be the ground state tetraquark state with the $J^{PC}=0^{++}$ or $1^{+-}$.

The predicted masses $M=6.57\pm0.09\,\rm{GeV}$ and $6.64\pm0.09\,\rm{GeV}$ for the 2S tetraquark states with the $J^{PC}=0^{++}$ and $1^{+-}$, respectively, are compatible with the values $6.62\pm0.03 {}^{+0.02}_{-0.01}\,\rm{GeV}$ from the ATLAS collaboration \cite{Atlas2022} and $6552\pm10\pm12 ~ \mathrm{MeV}$ from the CMS collaboration \cite{Cms2022}. And they  support assigning the $X(6620/6600)$ to be the first radial excited tetraquark state with the $J^{PC}=0^{++}$ or $1^{+-}$.
The predicted masses $M=6.52\pm0.10\,\rm{GeV}$ and $6.57\pm0.10\,\rm{GeV}$ for the $\tilde{V}\tilde{V}$-type ground state fully-charm tetraquark states with the  $J^{PC}=0^{++}$ and $1^{+-}$, respectively, in Ref.\cite{WZG-IJMPA-cccc}, are also compatible with the values $6.62\pm0.03 {}^{+0.02}_{-0.01}\,\rm{GeV}$ from the ATLAS collaboration \cite{Atlas2022} and $6552\pm10\pm12 ~ \mathrm{MeV}$ from the CMS collaboration \cite{Cms2022}. The assignments of the $X(6600/6620)$ as the ground state $\tilde{V}\tilde{V}$-type  tetraquark states with the $J^{PC}=0^{++}$ or $1^{+-}$ cannot be excluded.

 The predicted mass $M=6.92\pm0.09\,\rm{GeV}$ for the 3S tetraquark state with the $J^{PC}=0^{++}$ is compatible with the values $6.87\pm0.03 {}^{+0.06}_{-0.01}\,\rm{GeV}$ from the ATLAS collaboration \cite{Atlas2022} and $6927\pm9\pm5 ~ \mathrm{MeV}$ from the CMS collaboration \cite{Cms2022}. And it supports assigning the $X(6900)$ to be the second radial excited  tetraquark state with the $J^{PC}=0^{++}$.

 The predicted mass $M=7.25\pm0.09\,\rm{GeV}$ for the 4S tetraquark state with the $J^{PC}=0^{++}$ is compatible with the values $7.22\pm0.03 {}^{+0.02}_{-0.03}\,\rm{GeV}$ from the ATLAS collaboration \cite{Atlas2022} and $7287\pm19\pm5 ~ \mathrm{MeV}$ from the CMS collaboration \cite{Cms2022}. And it supports assigning the $X(7220/7300)$ to be the third radial excited  tetraquark state with the $J^{PC}=0^{++}$.

 In Table \ref{mass-cccc-Regge}, we list out the possible assignments as a summary. On the other hand, the ATLAS and CMS's new states can be generated dynamically and thus lead to possible molecule assignments according to the coupled channel effects \cite{GuoFK-2021-PRL,LiuX-2022-X7300}.

The LHCb  collaboration (also the ATLAS collaboration) also observed  some vague structures around
$7.2\,\rm{ GeV}$, which coincides with the $\overline{\Xi}_{cc}\Xi_{cc}$ threshold $7242.4\,\rm{MeV}$ \cite{LHCb-cccc-2006} (\cite{Atlas2022}).
  The energy is sufficient to create a $\overline{\Xi}_{cc}\Xi_{cc}$  pair  containing the valence quarks
$ccq \, \bar c \bar c \bar q$. In Ref.\cite{WZG-X7200}, we construct the color-singlet-color-singlet type six-quark pseudoscalar current to investigate
 the $\overline{\Xi}_{cc}\Xi_{cc}$ hexaquark molecular state with the QCD sum rules,
 the predicted mass $M_X \sim 7.2\,\rm{GeV}$ supports assigning the $X(7200)$ to be the $\overline{\Xi}_{cc}\Xi_{cc}$ hexaquark molecular state with the  $J^{PC}=0^{-+}$. The decay $X(7200)\to J/\psi J/\psi$ can take place through fusions of the $\bar{q}q$ pairs. Whether or not the $X(7200)$ and $X(7300)$ are the same particle needs further  experimental data.

\section{Conclusion}
In this work, we re-investigate  the mass spectrum of the ground state, first radial, second radial and third radial excited  diquark-antidiquark type fully-charm  tetraquark states with the QCD sum rules. We  take account of the new CMS and ATLAS experimental data and preform a self-consistent  analysis, and try to make possible assignments of the $X(6600)$, $X(6900)$ and $X(7300)$  consistently. We choose the uniform $\overline{MS}$ charm quark mass, self-consistent continuum threshold parameters and suitable energy scales of the QCD spectral densities to acquire stable QCD sum rules for the ground states and first radial excited states. Then we resort to the Regge trajectories to acquire the second and third radial excited states. The predictions support assigning  the $X(6220)$ to be the ground state tetraquark state with the $J^{PC}=0^{++}$ or $1^{+-}$, assigning the $X(6620/6600)$ to be the first radial excited  tetraquark state with the $J^{PC}=0^{++}$ or $1^{+-}$, assigning the $X(6900)$ to be the second radial excited tetraquark state with the $J^{PC}=0^{++}$, and assigning the $X(7220/7300)$ to be the third radial excited tetraquark state with the $J^{PC}=0^{++}$. More experimental data are still needed to acquire more robust assignments.

\section*{Appendix}
The QCD spectral densities,
\begin{eqnarray}
\rho_{S/A/V/T}(s)&=& \int_{z_i}^{z_f}dz \int_{t_i}^{t_f}dt \int_{r_i}^{r_f}dr\, \rho_{S/A/V/T}(s,z,t,r) \, ,
 \end{eqnarray}
 where
\begin{eqnarray}
\rho_S(s,z,t,r)&=& \frac{3m_c^4}{8\pi^6}\left( s-\overline{m}_c^2\right)^2+\frac{t z m_c^2}{8\pi^6}\left( s-\overline{m}_c^2\right)^2\left( 5s-2\overline{m}_c^2\right) \nonumber\\
&&+\frac{rtz(1-r-t-z)}{1-t-z} \frac{1}{32\pi^6}\left( s-\overline{m}_c^2\right)^3\left( 3s-\overline{m}_c^2\right) \nonumber\\
&&+\frac{rtz(1-r-t-z)}{1-z} \frac{1}{32\pi^6}\left( s-\overline{m}_c^2\right)^3\left( 3s-\overline{m}_c^2\right)\left[5-\frac{t}{1-t-z} \right] \nonumber\\
&&-\frac{rtz^2(1-r-t-z)}{1-z} \frac{3}{16\pi^6}\left( s-\overline{m}_c^2\right)^4 \nonumber\\
&&+rtz(1-r-t-z) \frac{3s}{8\pi^6}\left( s-\overline{m}_c^2\right)^2\left[ 2s-\overline{m}_c^2-\frac{z}{1-z}\left( s-\overline{m}_c^2\right)\right] \nonumber\\
&&+m_c^2\langle \frac{\alpha_sGG}{\pi}\rangle \left\{-\frac{1}{r^3} \frac{m_c^4}{6\pi^4}\delta\left( s-\overline{m}_c^2\right) -\frac{1-r-t-z}{r^2} \frac{m_c^2}{12\pi^4}\left[2+s\,\delta\left( s-\overline{m}_c^2\right)\right]\right. \nonumber\\
&&-\frac{tz}{r^3} \frac{m_c^2}{12\pi^4}\left[2+s\,\delta\left( s-\overline{m}_c^2\right)\right]  -\frac{tz(1-r-t-z)}{r^2(1-t-z)} \frac{1}{12\pi^6}\left( 3s-2\overline{m}_c^2\right) \nonumber\\
&&-\frac{tz(1-r-t-z)}{r^2(1-z)} \frac{1}{12\pi^4}\left( 3s-2\overline{m}_c^2\right) \left[5-\frac{t}{1-t-z} \right]\nonumber\\
&&+\frac{tz^2(1-r-t-z)}{r^2(1-z)} \frac{1}{\pi^4}\left( s-\overline{m}_c^2\right)  \nonumber\\
&&-\frac{tz(1-r-t-z)}{r^2} \frac{1}{2\pi^4}\left[s+\frac{s^2}{3}\delta\left( s-\overline{m}_c^2\right)-\frac{z}{1-z}s\right]  \nonumber\\
&&\left.+\frac{1}{r^2} \frac{m_c^2}{2\pi^4}   +\frac{tz}{r^2} \frac{1}{4\pi^4}\left( 3s-2\overline{m}_c^2\right)-  \frac{1}{16\pi^4}\left( 3s-2\overline{m}_c^2\right)  \right\} \nonumber
\end{eqnarray}
\begin{eqnarray}
&&+\langle \frac{\alpha_sGG}{\pi}\rangle \left\{\frac{1}{rz} \frac{m_c^4}{6\pi^4} +\frac{t}{r} \frac{m_c^2}{6\pi^4}\left( 3s-2\overline{m}_c^2\right)\right. \nonumber\\
&&+\frac{t(1-r-t-z)}{(1-t-z)} \frac{1}{12\pi^4}\left( s-\overline{m}_c^2\right) \left( 2s-\overline{m}_c^2\right)  \nonumber\\
&&+\frac{t(1-r-t-z)}{(1-z)} \frac{1}{12\pi^4}\left( s-\overline{m}_c^2\right) \left( 2s-\overline{m}_c^2\right) \left[2-\frac{t}{1-t-z} \right] \nonumber\\
&&-\frac{tz(1-r-t-z)}{(1-z)} \frac{1}{4\pi^4}\left( s-\overline{m}_c^2\right)^2\nonumber\\
&&\left.+t(1-r-t-z) \frac{1}{12\pi^4}s\left[ 4s-3\overline{m}_c^2-\frac{z}{1-z}3\left( s-\overline{m}_c^2\right)\right]\right\}\, ,
\end{eqnarray}

\begin{eqnarray}
\rho_T(s,z,t,r)&=& \frac{3m_c^4}{16\pi^6}\left( s-\overline{m}_c^2\right)^2+\frac{t z m_c^2}{8\pi^6}\left( s-\overline{m}_c^2\right)^2\left( 4s-\overline{m}_c^2\right) \nonumber\\
&&+\frac{rtz(1-r-t-z)}{1-t-z} \frac{1}{320\pi^6}\left( s-\overline{m}_c^2\right)^3\left( 17s-5\overline{m}_c^2\right) \nonumber\\
&&+\frac{rtz(1-r-t-z)}{1-z} \frac{1}{320\pi^6}\left( s-\overline{m}_c^2\right)^3\left[\left( 21s-5\overline{m}_c^2\right)-\frac{t}{1-t-z}\left( 17s-5\overline{m}_c^2\right) \right] \nonumber\\
&&-\frac{rtz^2(1-r-t-z)}{1-z} \frac{1}{32\pi^6}\left( s-\overline{m}_c^2\right)^4 \nonumber\\
&&+rtz(1-r-t-z) \frac{s}{80\pi^6}\left( s-\overline{m}_c^2\right)^2\left[ 28s-13\overline{m}_c^2-\frac{z}{1-z}7\left( s-\overline{m}_c^2\right)\right] \nonumber\\
&&+m_c^2\langle \frac{\alpha_sGG}{\pi}\rangle \left\{-\frac{1}{r^3} \frac{m_c^4}{12\pi^4}\delta\left( s-\overline{m}_c^2\right) -\frac{1-r-t-z}{r^2} \frac{m_c^2}{12\pi^4}\left[1+s\,\delta\left( s-\overline{m}_c^2\right)\right]\right. \nonumber\\
&&-\frac{tz}{r^3} \frac{m_c^2}{12\pi^4}\left[1+s\,\delta\left( s-\overline{m}_c^2\right)\right]  -\frac{tz(1-r-t-z)}{r^2(1-t-z)} \frac{1}{12\pi^6}\left( 2s-\overline{m}_c^2\right) \nonumber\\
&&-\frac{tz(1-r-t-z)}{r^2(1-z)} \frac{1}{12\pi^4}\left( 2s-\overline{m}_c^2\right) \left[1-\frac{t}{1-t-z} \right]\nonumber\\
&&+\frac{tz^2(1-r-t-z)}{r^2(1-z)} \frac{1}{6\pi^4}\left( s-\overline{m}_c^2\right)  \nonumber\\
&&-\frac{tz(1-r-t-z)}{r^2} \frac{1}{6\pi^4}\left[s+\frac{s^2}{2}\delta\left( s-\overline{m}_c^2\right)-\frac{z}{1-z}s\right]  \nonumber\\
&&\left.+\frac{1}{r^2} \frac{m_c^2}{4\pi^4}   +\frac{tz}{r^2} \frac{1}{4\pi^4}\left( 2s-\overline{m}_c^2\right)  \right\} \nonumber
\end{eqnarray}
\begin{eqnarray}
&&+\langle \frac{\alpha_sGG}{\pi}\rangle \left\{-  \frac{m_c^2}{48\pi^4}\left( 4s-3\overline{m}_c^2\right)\right. \nonumber\\
&&-\frac{r(1-r-t-z)}{1-t-z}\frac{1}{32\pi^4} \left( s-\overline{m}_c^2\right)\left( 3s-\overline{m}_c^2\right)\nonumber\\
&&-\frac{r(1-r-t-z)}{1-z}\frac{1}{480\pi^4} \left( s-\overline{m}_c^2\right)\left[\left( 17s-5\overline{m}_c^2\right)-\frac{t}{1-t-z}15\left( 3s-\overline{m}_c^2\right)\right]\nonumber\\
&&+\frac{rz(1-r-t-z)}{1-z}\frac{1}{24\pi^4} \left( s-\overline{m}_c^2\right)^2 \nonumber\\
&&-r(1-r-t-z)\frac{1}{240\pi^4} s\left[\left( 14s-9\overline{m}_c^2\right)-\frac{z}{1-z}21\left( s-\overline{m}_c^2\right)\right]\nonumber\\
&&-\frac{1}{rz} \frac{m_c^4}{36\pi^4} -\frac{t}{r} \frac{m_c^2}{18\pi^4}\left( 2s-\overline{m}_c^2\right)\nonumber\\
&&-\frac{t(1-r-t-z)}{(1-t-z)} \frac{1}{72\pi^4}\left( s-\overline{m}_c^2\right) \left( 4s-\overline{m}_c^2\right)  \nonumber\\
&&-\frac{t(1-r-t-z)}{(1-z)} \frac{1}{72\pi^4}\left( s-\overline{m}_c^2\right)  \left[2\left( 2s-\overline{m}_c^2\right)-\frac{t}{1-t-z}\left( 4s-\overline{m}_c^2\right) \right] \nonumber\\
&&+\frac{tz(1-r-t-z)}{(1-z)} \frac{1}{24\pi^4}\left( s-\overline{m}_c^2\right)^2\nonumber\\
&&\left.-t(1-r-t-z) \frac{1}{72\pi^4}s\left[ 7s-5\overline{m}_c^2-\frac{z}{1-z}5\left( s-\overline{m}_c^2\right)\right]\right\}\, ,
\end{eqnarray}

\begin{eqnarray}
\rho_A(s,z,t,r)&=& \frac{3m_c^4}{16\pi^6}\left( s-\overline{m}_c^2\right)^2+\frac{t z m_c^2}{8\pi^6}\left( s-\overline{m}_c^2\right)^2\left( 4s-\overline{m}_c^2\right) \nonumber\\
&&+rtz(1-r-t-z) \frac{s}{16\pi^6}\left( s-\overline{m}_c^2\right)^2\left( 7s-4\overline{m}_c^2\right) \nonumber\\
&&+m_c^2\langle \frac{\alpha_sGG}{\pi}\rangle \left\{-\frac{1}{r^3} \frac{m_c^4}{12\pi^4}\delta\left( s-\overline{m}_c^2\right) -\frac{1-r-t-z}{r^2} \frac{m_c^2}{12\pi^4}\left[1+s\,\delta\left( s-\overline{m}_c^2\right)\right]\right. \nonumber\\
&&-\frac{tz}{r^3} \frac{m_c^2}{12\pi^4}\left[1+s\,\delta\left( s-\overline{m}_c^2\right)\right] -\frac{tz(1-r-t-z)}{r^2} \frac{1}{12\pi^4}\left[4s+s^2\delta\left( s-\overline{m}_c^2\right)\right]  \nonumber\\
&&\left.+\frac{1}{r^2} \frac{m_c^2}{4\pi^4}   +\frac{tz}{r^2} \frac{1}{4\pi^4}\left( 2s-\overline{m}_c^2\right) \right\} \nonumber\\
&&+\langle \frac{\alpha_sGG}{\pi}\rangle \left\{-\frac{m_c^2}{48\pi^4}\left( 4s-3\overline{m}_c^2\right)- \frac{r(1-r-t-z)}{16\pi^4}\left( s-\overline{m}_c^2\right)^2\right. \nonumber\\
&&- \frac{r(1-r-t-z)}{48\pi^4}s\left( 7s-6\overline{m}_c^2\right)+\frac{1}{rz} \frac{m_c^4}{48\pi^4} +\frac{t}{r} \frac{m_c^2}{24\pi^4}\left( 2s-\overline{m}_c^2\right)\nonumber\\
&&\left.+ \frac{t(1-r-t-z)}{32\pi^4}\left( s-\overline{m}_c^2\right)^2+ \frac{t(1-r-t-z)}{48\pi^4}s\left( 6s-5\overline{m}_c^2\right)\right\}\, ,
\end{eqnarray}

\begin{eqnarray}
\rho_V(s,z,t,r)&=& -\frac{3m_c^4}{16\pi^6}\left( s-\overline{m}_c^2\right)^2-\frac{t z m_c^2}{8\pi^6}\left( s-\overline{m}_c^2\right)^3 \nonumber\\
&&+rtz(1-r-t-z) \frac{s}{16\pi^6}\left( s-\overline{m}_c^2\right)^2\left( 7s-4\overline{m}_c^2\right) \nonumber\\
&&+m_c^2\langle \frac{\alpha_sGG}{\pi}\rangle \left\{\frac{1}{r^3} \frac{m_c^4}{12\pi^4}\delta\left( s-\overline{m}_c^2\right)+\frac{1-r-t-z}{r^2} \frac{m_c^2}{12\pi^4}\right. \nonumber\\
&&+\frac{tz}{r^3} \frac{m_c^2}{12\pi^4} -\frac{tz(1-r-t-z)}{r^2} \frac{1}{12\pi^4}\left[4s+s^2\delta\left( s-\overline{m}_c^2\right)\right]  \nonumber\\
&&\left.-\frac{1}{r^2} \frac{m_c^2}{4\pi^4}   -\frac{tz}{r^2} \frac{1}{4\pi^4}\left( s-\overline{m}_c^2\right) \right\} \nonumber\\
&&+\langle \frac{\alpha_sGG}{\pi}\rangle \left\{\frac{m_c^2}{48\pi^4}\left( 5s-3\overline{m}_c^2\right)+ \frac{r(1-r-t-z)}{16\pi^4}\left( s-\overline{m}_c^2\right)^2\right. \nonumber\\
&&+ \frac{r(1-r-t-z)}{48\pi^4}s\left( 7s-6\overline{m}_c^2\right)-\frac{1}{rz} \frac{m_c^4}{48\pi^4} -\frac{t}{r} \frac{m_c^2}{24\pi^4}\left( s-\overline{m}_c^2\right)\nonumber\\
&&\left.- \frac{t(1-r-t-z)}{32\pi^4}\left( s-\overline{m}_c^2\right)^2- \frac{t(1-r-t-z)}{48\pi^4}s\left( s-\overline{m}_c^2\right)\right\}\, ,
\end{eqnarray}
and
\begin{eqnarray}
\overline{m}_c^2&=&\frac{m_c^2}{r}+\frac{m_c^2}{t}+\frac{m_c^2}{z}+\frac{m_c^2}{1-r-t-z}\, ,\nonumber
\end{eqnarray}
\begin{eqnarray}
r_{f/i}&=&\frac{1}{2}\left\{1-z-t \pm \sqrt{(1-z-t)^2-4\frac{1-z-t}{\hat{s}-\frac{1}{z}-\frac{1}{t}}}\right\} \, ,\nonumber\\
t_{f/i}&=&\frac{1}{2\left( \hat{s}-\frac{1}{z}\right)}\left\{ (1-z)\left( \hat{s}-\frac{1}{z}\right)-3 \pm \sqrt{ \left[  (1-z)\left( \hat{s}-\frac{1}{z}\right)-3\right]^2-4 (1-z)\left( \hat{s}-\frac{1}{z}\right)  }\right\}\, ,\nonumber\\
z_{f/i}&=&\frac{1}{2\hat{s}}\left\{ \hat{s}-8 \pm \sqrt{\left(\hat{s}-8\right)^2-4\hat{s}  }\right\}\, ,
\end{eqnarray}
and $\hat{s}=\frac{s}{m_c^2}$.

\section*{Acknowledgements}
This  work is supported by National Natural Science Foundation, Grant Number  12175068.


\begin{thebibliography}{99}

\bibitem{LHCb-cccc-2006} R. Aaij et al, Sci. Bull. {\bf 65} (2020) 1983.

\bibitem{Atlas2022}
E. Bouhova-Thacker on behalf of the ATLAS Collaboration,
ATLAS results on exotic hadronic resonances, Proceedings at ICHEP 2022,\\
https://agenda.infn.it/event/28874/contributions/170298/.

\bibitem{Cms2022}
K. Yi on behalf of the CMS Collaboration,
Recent CMS results on exotic resonance, Proceedings at ICHEP 2022, \\
https://agenda.infn.it/event/28874/contributions/170300/.


\bibitem{Rosner-2017} M. Karliner, J. L. Rosner and S. Nussinov,  Phys. Rev. {\bf D95} (2017)  034011.

\bibitem{WZG-QQQQ-1} Z. G. Wang, Eur. Phys. J. {\bf C77} (2017) 432.


\bibitem{Chen-2017}  W. Chen, H. X. Chen, X. Liu, T. G. Steele and S. L. Zhu, Phys. Lett. {\bf B773} (2017) 247.

\bibitem{Wu-2018} J. Wu, Y. R. Liu, K. Chen, X. Liu and S. L. Zhu, Phys. Rev. {\bf D97} (2018) 094015.

\bibitem{FKGuo-2018-Anwar} M. N. Anwar, J. Ferretti, F. K. Guo, E. Santopinto and B. S. Zou, Eur. Phys. J. {\bf C78} (2018)  647.

\bibitem{Polosa-2018} A. Esposito and A. D. Polosa, Eur. Phys. J. {\bf C78} (2018)  782.

\bibitem{Hughes-2018}  C. Hughes, E. Eichten and C. T. H. Davies, Phys. Rev. {\bf D97} (2018)  054505.


\bibitem{Navarra-2019}    V. R. Debastiani and F. S. Navarra, Chin. Phys. {\bf C43} (2019)  013105.

\bibitem{Bai-2016}  Y. Bai, S. Lu and J. Osborne, 	Phys. Lett. {\bf B798} (2019) 134930.

\bibitem{WZG-QQQQ-2} Z. G. Wang and Z. Y. Di,  Acta Phys. Polon. {\bf B50} (2019) 1335.

\bibitem{Roberts-2020} M. A. Bedolla, J. Ferretti, C. D. Roberts, and E. Santopinto,   Eur. Phys. J. {\bf C80} (2020) 1004.

\bibitem{WZG-CPC-cccc} Z. G. Wang, Chin. Phys. {\bf C44} (2020)  113106.


\bibitem{PingJL-2020} X. Jin, Y. Xue, H. Huang and J. Ping,	Eur. Phys. J. {\bf C80} (2020) 1083.

\bibitem{Zhong-2019} M. S. Liu, Q. F. Lu, X. H. Zhong and Q. Zhao, Phys. Rev. {\bf D100} (2019)  016006.

\bibitem{WangJZ-Produ-mass} J. Z. Wang, D. Y. Chen, X. Liu, and T. Matsuki,	Phys. Rev. {\bf D103} (2021) 071503.

\bibitem{Zhong-2021} F. X. Liu, M. S. Liu, X. H. Zhong and Q. Zhao, Phys. Rev. {\bf D104} (2021) 116029.

\bibitem{Zhu-2021-NPB} R. Zhu,	Nucl. Phys. {\bf B966} (2021) 115393.

\bibitem{WZG-IJMPA-cccc} Z. G. Wang, Int. J. Mod. Phys. {\bf A36} (2021)  2150014.

\bibitem{ZhangJR-PRD} J. R. Zhang, Phys. Rev. {\bf D103} (2021)  014018.

\bibitem{QiaoCF-2021} B. C. Yang, L. Tang, and C. F. Qiao,	Eur. Phys. J. {\bf C81} (2021) 324.

\bibitem{GuoFK-2021-PRL} X. K. Dong, V. Baru, F. K. Guo, C. Hanhart and A. Nefediev, Phys. Rev. Lett. {\bf 127} (2021) 119901.

\bibitem{XKDong-SB-2021} X. K. Dong, V. Baru, F. K. Guo, C. Hanhart and A. Nefediev, Sci. Bull. {\bf 66} (2021)  2462.


\bibitem{Gong-Zhong-2022} C. Gong, M. C. Du, Q. Zhao, X. H. Zhong and B. Zhou, Phys. Lett. {\bf B824} (2022) 136794.

\bibitem{Baxi-G} M. S. Maior de Sousa and R. Rodrigues da Silva, Braz. J. Phys. {\bf 46} (2016) 730.

\bibitem{WangZG-4430-1} Z. G. Wang, Commun. Theor. Phys. {\bf 63} (2015)  325.

\bibitem{WangZG-4430-2} Z. G. Wang, Chin. Phys. {\bf C44} (2020)  063105.

\bibitem{SVZ79-1}  M. A. Shifman, A. I. Vainshtein and V. I. Zakharov, Nucl. Phys. {\bf B147} (1979) 385.

\bibitem{SVZ79-2}  M. A. Shifman, A. I. Vainshtein and V. I. Zakharov, Nucl. Phys. {\bf B147} (1979) 448.

\bibitem{Reinders85} L. J. Reinders, H. Rubinstein and S. Yazaki, Phys. Rept. {\bf 127} (1985) 1.


\bibitem{ColangeloReview} P. Colangelo and A. Khodjamirian, hep-ph/0010175.

\bibitem{PDG}  P. A. Zyla et al,  Prog. Theor. Exp. Phys. {\bf 2020} (2020) 083C01.

\bibitem{WZG-formula-mole} Z. G. Wang, Eur. Phys. J. {\bf C74} (2014)  2963.

\bibitem{WZG-formula} Z. G. Wang, Eur. Phys. J. {\bf C74} (2014) 2874.

 \bibitem{WZG-PRD-hidden-charm} Z. G. Wang, Phys. Rev. {\bf D102} (2020)  014018.

\bibitem{WZG-CPC-Zcs3985} Z. G. Wang, Chin. Phys. {\bf C45} (2021)  073107.

\bibitem{WZG-P4312-IJMPA} Z. G. Wang, Int. J. Mod. Phys. {\bf A35} (2020)  2050003.

\bibitem{WZG-Pcs4459-IJMPA} Z. G. Wang, Int. J. Mod. Phys. {\bf A36} (2021)  2150071.

\bibitem{LiuX-2022-X7300} J. Z. Wang and X. Liu, arXiv: 2207.04893 [hep-ph].


\bibitem{WZG-X7200} Z. G. Wang,  Phys. Lett. {\bf B819} (2021) 136464.


\end{thebibliography}
\end{document}